**Could the number of blue straggler stars help to determine the age of their parent globular cluster?**

Félix Llorente de Andrés[1,2],
[1] Centro de Astrobiología (CAB), CSIC-INTA, Camino Bajo del Castillo s/n, campus ESAC, 28692, Villanueva de la Cañada, Madrid, Spain.
[2] Ateneo de Almagro, Sección de Ciencia y Tecnología, 13270 Almagro, Spain

**ABSTRACT**

A recent study shows, from an empirical deduction, that the number and the presence of the blue straggler stars (BSS) in an open cluster follow a function whose components are the ratio between age and the relaxation time, ƒ, and a factor, ϖ, which is an indicator of stellar collisions plus primordial binaries. The relation among the number of blue straggler stars, the factor ƒ and the factor ϖ of each globular cluster allows for deriving the age of the respective globular clusters. This method has been applied individually over 56 globular clusters containing BSS. The values derived for the cluster ages from our methodology do not differ from those derived from other methods. A special case is the cluster NGC104 whose age exceeds 13.8 Gyr (its age is in between 19.04 and 20.30Gyr), which would have a very exotic explanation: the existence of an intermediate black hole in the center of the cluster. That black hole main-sequence star (BH-MS) binaries with an initial orbital period less than the bifurcation period can evolve into ultra-compact X-ray binaries (UCXBs) that can be detected by LISA. On the other hand, if that age were true, it would call into question the expansion velocity for a flat Universe. This would call into question the case for a dark energy–dominated Universe.

**INTRODUCTION**

The determination of the age of the oldest clusters is so important insomuch as of these ages becoming the most stringent lower limits on the age of our galaxy, and the Universe as well. However stellar models and the methods for the age determinations of globular clusters are still in need of improvement.

There are three independent ways to reliably infer the age of the oldest stars in our galaxy: nucleocosmochrology, white dwarf cooling curves, and the main sequence turnoff time scale.

The first way is based on the Thorium ($^{232}$Th) and Uranium ($^{238}$U) abundance measurements and a comparison of these abundances to the abundance of other r-process elements. This method presents difficulties because Th and U have weak spectral lines so this can only be done with stars with enhanced Th and U (requires large surveys for metal-poor stars) and unknown theoretical predictions for the production of r-process (rapid neutron capture) elements. Based on this method, an age of 13.84 ± 4 Gyr is found for the star metal-poor star BD +17 3248 (J. J. Cowan et al., 2002). For CS 31082-001, the Uranium abundance alone gives an age of 12.5 ± 3 Gyr while the Th/Ur ratio of this same star gives 14.0 ± 2.4 Gyr (Cayrel et al. 2001). These ages are consistent with those obtained by other methods. The second method, white dwarf cooling curves, was performed by Hense et al (2004) based on a deeper exposure of NGC 6121 with HST (Hubble Space Telescope). They estimate age for this cluster of 12.7± 0.7 Gyr. This age is compatible with ages found for other globular clusters by means of another method: the main sequence turn-off. The third method, main sequence turnoff ages, which has the smallest theoretical uncertainties; the method is based on the luminosity at this "main sequence turnoff" point.

However it has been found that the uncertainty in translating magnitudes to luminosities (i.e., the uncertainties in deriving distances to globular clusters, such inaccuracy is expected to be surpassed by GAIA) is the main source of uncertainty in globular cluster age estimates. Some examples would be: Gratton et al (2003) find for NGC 6397 & NGC 6752 ages of 13.9 ± 1.1 and 13.8 ±1.1 Gyr respectively. For 47 Tuc (NGC 104)the age would be 11.3 ±1.1 Gyr and 13.4 ± 0.8 Gyr for the oldest clusters.

Nevertheless, there are another methods as the recent study performed by Florentino et al (2016) combining the Cosmic Microwave Background and Baryonic Acoustic Oscillations and the method of turn off (TO) found age values for NGC 2808 of 10.9 ± 7 Gyr. Bono et al (2010), based on the difference in magnitude between the main-sequence turnoff (MSTO) and the lower main sequence (MSK), estimate for NGC 3201 and age estimates 11.48 ± 1.27 and 11.55 ± 1.53 Gyr. Massari et al (2016) using deep IR found that NGC 2808 has an age of 10.9 ± 0.7 (intrinsic) ±0.45 (metallicity term). And just for finalizing the frame, Correnti et al (2016) from the Hubble Space Telescope Wide Field Camera 3 IR archival observations of four GCs—47 Tuc (NGC 6626), M4 (NGC 6121), NGC 2808, and NGC 6752—for which they derived the fiducial lines for each cluster and compared them with a grid of isochrones over a large range of parameter space, allowing age, metallicity, distance, and reddening to vary within reasonable selected ranges. The derived ages for the four clusters are, respectively, 11.6, 11.5, 11.2, and 12.1 Gyr.

If all these results are used to constrain cosmological parameters, one needs to add to these ages a time that corresponds to the time between the Big Bang and the formation of globular clusters in our galaxy. Let it be supposed that the age of the Universe is 13.8 Gyr, margin 20%, which means: the oldest clusters should be younger than 13.6 Gyr; this figure becomes the upper limit

Interestingly, for the best fit value of the Hubble constant, globular cluster age limits also put strong limits on the total matter density of the Universe; but it is no matter of this work.

This work attempts to obtain a more objective method of age determination based on the abundance of blue straggler stars (BSS). This way of deriving globular cluster ages is attempted for the first time. It is not an irrational idea because derived ages based on the turn-off are worthy ages. Why not a method based on the number of BSS wouldn't work.

This work tries to obtain new method of age determinations. This method has, as the others do, some inconsistencies and is not trouble free. Anyway, the method contributes to obtain very consistent ages.

In a previous work (Llorente de Andrés and Morales Durán, 2022 here in after LM), was explained that the presence of BSS in a cluster is not accidental, nor random, but its discrete number follows a function which is related to the ratio between age of the cluster and relax time (called ƒ = age/trelax) and to a factor, we call $\varpi$, which is the indicator of stellar collisions plus primordial binary evolution thus related to the total number of observed BSS (NBS). The number of BSS follows the function: $NBS \cong f^3 \left( \frac{1}{e^{f/\varpi}-1} \right)$, called equation 6 of the referenced paper and explained in Section 2. This function was applied to a complete set of observed number of BSS of globular clusters (GC) listed by Moretti et al (2008), hereinafter MAP; this number of BSS is displayed within MAP's paper in the table, fourth column under the label NBS,tot. LM calculated the number of theoretical BSS by assuming that the relax time might be the value alike to that one listed as the median relaxation time t(r_h) published by Harris (1996; but edition 2010),herein after H10. As GC median age it was adopted the value suggested by Kraus and Chaboyer (2003), herein after KC: 12.6 Gyr. Thus, it was computed, cluster by cluster, by means of the above mentioned equation the number of predicted BSS versus the number of observed BSS; the relation was found one to one with an $R^2 \cong 1$.

In order to extend the data, not restricted to only one, catalogue this work was also carried out with another BSS catalogue published by Leigh et al (2007), herein after LSK. As well as to avoid bias because of a single source, the present study was also performed by taking the median relaxation time from Recio-Blanco et al (2006), herein after RB.

It is also determined the goodness of results by comparing them to different sets of age published by other authors, as those collected by Meissner and Weiss (2006), herein after MW, who included their own values, and the values published by VandenBerg et al (2013), herein after VBLC. Such comparison give an estimate of the confidence level of the present results within the uncertainties, not always accessible to reach ; values quoted from MW and VBLC and those derived from the present work are displayed in Table 5; including a proposal for an age minimum and age Maximum in the last two columns.

In addition, it can be affirmed that the metallicity does not play any relevant role.

This work is divided into the following sections: the model and its methodology and its results are introduced and discussed in Section 2 along with particular details of fractious clusters because of erroneous age. The presentation and discussion of results are displayed in Section 3 with references to previous works by comparison of the present work calculations face to those published by other authors. Section 4 is devoted to the conclusions.

Something is very clear: the determination of globular cluster ages still depends on the favorite method or indicator chosen.

There was the following sentence written in the LM's article which launched the present work: "The simple though that equation (6) could possibly help in determining the maximum and minimum value of GC ages is overwhelming".

## 2. MODEL AND ITS METHODOLOGY

### 2.1 The model

The aim of this section is to explain the model basis of the present study.

In a previous paper LM showed that the number of presence of BSS could be computed by means of the equation, identified in this paper as equation 6: $NBS \cong f^3 \left( \frac{1}{e^{f/\varpi}-1} \right)$ . Where ƒ= age/trlx , which is an indicator of the probability of encounter and/or deflection of orbits. This ratio ƒ is implicitly normalized to the total number of stars, N; thus ƒ is able to be compared among clusters, irrespective of their mass. The factor $\varpi$ is, in fact, the indicator of stellar collisions plus primordial binary evolution related to the total number of observed BS stars.

It was verified whether such LM's equation 6, reproduces the number of BS stars present in a cluster by means of an empirical auditory. That was carried out cluster by cluster, for the two samples of open clusters (Ahumada and Lapasset (2007), de Marchi et al. (2014) and for one of globular clusters (MAP) – in thepresent study it is included the catalogue published by LSK - showing in every case that the number of observed BS stars which appears in OC and GC is not accidental, nor random, but rather based on a law as expressed and formulated, by the above mentioned equation.

This mentioned equation is the origin of the model which is used in the present work to derive the age of the two sets of GC; whose content of BSS has been published in MAP's and LSK's catalogues. The intention is to have a spreader number of BSS quoted from different catalogues; in this case the difference between the two used catalogues (MAP and LSK) comes from the different normalization applied in each identification and collection of BSS. MAP normalize absolute number of BSS in a region divided by the total luminosity of stars in the same region. However LSK normalize the BSS absolute number to different cluster populations. In the first case the number of BSS is higher than in the second case but that helps to have references in the present study which will be very useful for stablishing a minimum and a maximum for the GC ages of this study sample. Let it be remembered before going further that mentioned equation, in the present study, was computed assuming two median relaxation times published by H10 and by RB; and an adopted median age for GCs, as suggested by KC, as 12.6 Gyr.

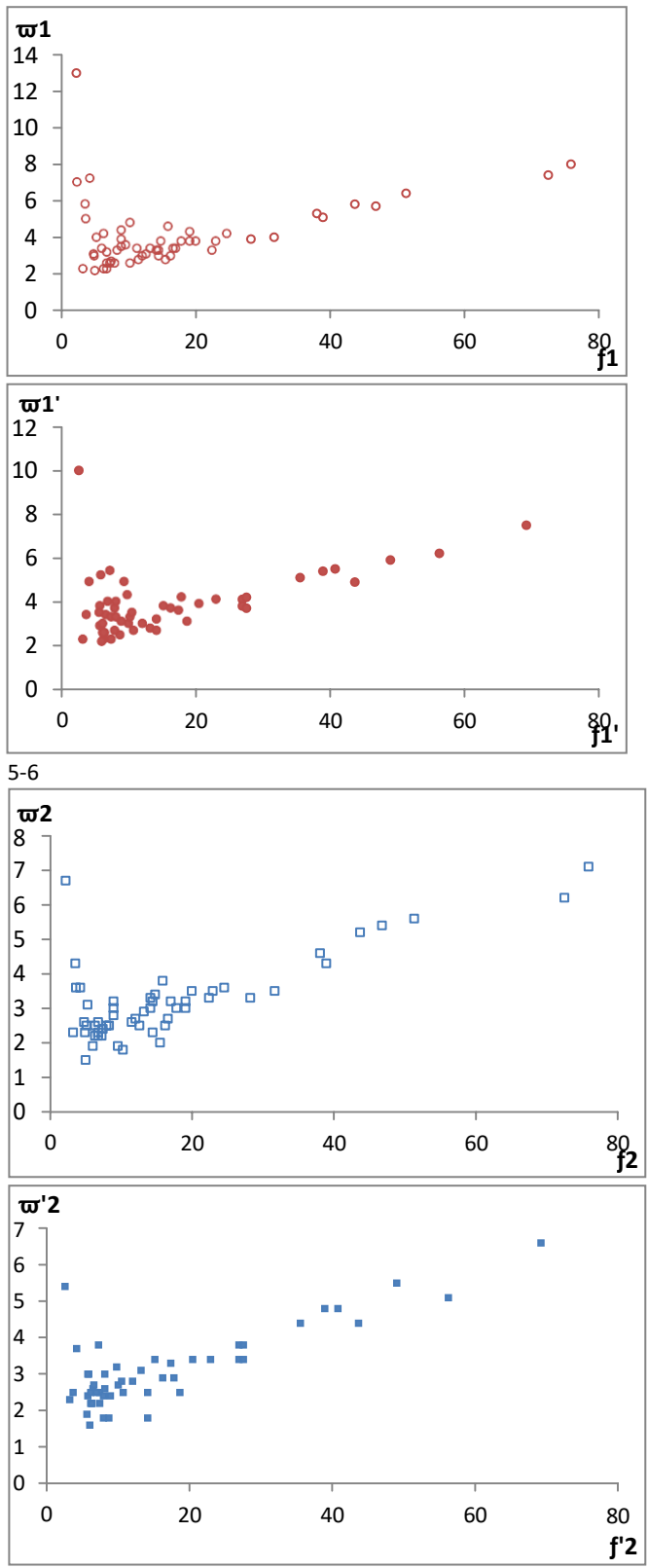

Fig. 1 shows the relationships between ϖ and ƒ for the three cases.
Top: factors derived from MAP catalogue and H10 trlx. Midlle top: factors derived from MAP catalogue and RB trlx. Midlle bottom: factors derived from LSK catalogue and H10 trlx.
Bottom: factors derived from LSK catalogue and H10 trlx.

In the four graphics it is easy to distinguish two separate regions; ƒ < 6 and ϖ > 3; the difference in the last value is due to the different number of BSS listed in both catalogues.

As the, so much mentioned, equation relates the number of BSS in a GC with the factor ϖ, with the trlx and with the age – the two last variables through the factor ƒ -, it should be easy to find a way to derive the corresponding age for the GC sample. This way is found by means the relationships between [ƒ], [ϖ] – median of ƒ and ϖ, respectively – published by LM. The Fig. 1 shows this relationship for the four cases. It is clear that the relation ϖ versus ƒ is well represented by a straight line; except the region where ƒ is smaller than ~ 6 and ϖ upper than ~3; thus this stage will have a separate treatment. The methodology is explained in the following Subsection 2.2.

**2.2 The methodology and the results**

The process is a reiterative process. Firstly, let it be found a correlation ϖ vs ƒ which shall be a linear relation. Thus look for an equation (ϖ = A*ƒ + B), which exhibits the best fit; a new value of ƒ, named $f_T$, is then derived as a function of ϖ. This value $f_T$ allows to derive the GC age (age = $f_T$ * trlx). The process starts at an exceptional stage: this part of the graphic which shows a big dispersion (see fig 1). These values are put aside and the process continues with the rest of the values.

However at this stage it is necessary to take into account a cut off because it is assuming that the globular clusters are not older than 13.6 Gyr which is compatible with an age for the Universe of ~ 13.8 Gyr; meaning that those clusters whose computed ages are higher than 13.6 Gyr are treated separately. The process is repeated, with the rest of the values. At each step, a new equation is solved then a new $f_T$ is obtained yielding to a new value for age. In the same way is done if the value of calculated age is negative. All values are displayed in Table 1, Table 2, Table 3 and Table 4. The tables are reproduced after the references (not disturbing the continuous lecture).

All these tables are structured in the same way: the first row shows both the stage and the transform equation - including the $R^2$ - or information about the calculi. The columns are respectively: cluster name, adopted trelax, primitive ƒ, primitive ϖ, theoretical number of BSS, observed number of BSS, $f_T$ which is the value of ƒ derived from the linear equation and, finally, the theoretical age proposed for the respective cluster.

A bird eye view shows that: those clusters with negative ages (erroneous) exhibit a very low number of BSS combined with a very low $f_T$ values as well MAP as LSK catalogues. By the contrary, those clusters with very high age (> 13.6 Gy) have low $f_T$ values but large values of the number of BSS; in both BSS catalogues – reminder: these clusters were deserved a s e p a r a t e study. A common cluster with wrong age value is NGC 7089 (13 Gyr in the Messier catalogue). Kuzman et al (2016) affirms that this cluster is diluting, hosting stellar populations with a broad dispersion in abundances and variety of neutron capture; they suggest this cluster as a stripper nucleus of a dwarf galaxy. Some other wrong age cases could be explained by different arguments. For instance: Muñoz et al (2013) give to NGC 3201 an extragalactic origin. The error of the age computation for NGC 4372 could come for the fact that there is an unusually high ratio of rotation amplitude to velocity dispersion what suggests a flattening which is indeed caused by the systemic rotation rather than tidal interactions with the Galaxy (Kacharov et al, 2014), etc.

In spite of the abovementioned divergence respect to the computed ages, some of the ages of these clusters diverge from those values computed by other authors for instance: NGC 104 (too high values of age), NGC 5024 (too low values of ages), NGC 5986 (low values of ages), NGC 6284 (too low values of ages) and NGC 7099 (too low values of ages).

NGC 5024 is kwon as an FG (first generation) because it might be composed primarily of first generation stars (Boberg et al (2016); the consequence is an erroneous value of the number of BSS, consequently a wrong value of $\varpi$. The low value of age of NGC 5986 could be due to an error either in the number of BSS (50 in MAP and 21 in LSK). It is also suggested for NGC 6284 (22 in MAP and 4 in LSK). In the case of NGC 7099 close encounters between normal stars and collapsed stars could mask values of the trelax.

A very interesting case is NGC 104, the correction could be done by increasing the value of the relaxation time which is feasible according to the suggestion of Krause et al (2016) or the possible existence to a more exotic proposal: the existence of an intermediate black hole in the center of the cluster (Kizilton et al 2017). Recently Ke Qin et al 2023 simulations showed that black hole main-sequence star (BH-MS) binaries with an initial orbital period less than the bifurcation period can evolve into ultra-compact X-ray binaries (UCXBs) that can be detected by LISA.

All previous ages calculated in this work and those proposed by other authors have been put together to create a unique table (Table 5) The fact of listing these values jointly facilitates the comparison among different authors. The table 5 lists: first column cluster name; second and third columns: metallicity from H10, metallicity from M&W and from VBLC, respectively; fifth to twelfth columns: ages (in Gyr) for the clusters quoted from M&S; thirteenth column ages (in Gyr) for the clusters, containing BSS, quoted from VBLC; fourteenth to sixteenth columns the ages (in Gyr) computed with the model of the present work taking the number of BSS from MAP and from LSK and adopting trlx from H10 and RB (each identification is described on the second row of the column); and finally the last two columns displayed the proposed minimum and Maximum age values for the globular clusters, studied in this work. These minimum and Maximum values of age for the GC of the present study are compatible with the rest of authors who computed them through different methods.

### 3. Presentation of results

It is tantalizing to compare age values computed by different authors who made them from different, and sometimes, no related methods, the question is: how does it do? Because as it is generally accepted that the error in determining values of age is very high. Let it be firstly compared the median values (listed in the bottom row of table 5). In the present work the median ages, for the minimum and the Maximum values, vary from 7.6 to 12. Gyr. and, for the same sample the median values but published by M&W and by VBLC vary from 8.8 to 12 Gyr. The model and the methodology drive to obtain reliable age values for our sample of GC; at least as reliable as the other authors reached by means of different models.

Herewith the already mentioned Table 5 is displayed with a complete set of age values, including those found in this study. Let it be the reader and/or future user who catches the value which is the more reliable up to him. In order to facilitate his selection it has been marked in slight dark those age values generating the major doubts. As the computed values of age in this model depends on the number of BSS, the factor $\varpi$ and the trlx – through the factor $f_T$ – all those values which are in a doubt might be reconsidered by means of these three variables but mostly due to the meaning of $\varpi$ and the correlation between the two sources of the relaxation time (trlx). There is a linear relation between them: logtrlx RB = 0.9 logtrlx H+ 0.95 with an $R^2$ = 0.9, as depicted in Figure 4. The big dispersion is found at the values around logtrlx ~ 9; such reason is valid for some extreme cases NGC1261, NGC 5946, NGC 6723, NGC 7078.

The real reason for those "possible wrong" values has to be found in number of BSS; moreover there is big difference on the content of number of BSS from the two catalogues Figure 5 shows up this difference. Just for having a roughly idea, the average of the ratio between them (nbr of BSSMAP/ nbr of BSSLSK) is around 3 times (in some clusters the ratio is higher than 6 even in more than 11 times). Thus these inconsistences come from the fact of using different catalogues which differ in the way of determining the number of BSS that leads to a different number of BSS (higher in the case of MAP; their proposed number of BSS is overestimate?) and additionally the way of normalization biases the number of BSS.

### 4. Discussion and conclusions

The aim of this work is to present a new method of age determinations. Although the method has, as the others do, some inconsistences and is not trouble free it contributes to obtain a new list of very consistent ages

The method is based on the results from a previous work, LM, where it was showed that the presence of BSS in a cluster is not accidental, nor random; the number of BSS a function whose variables are: the ratio between age of the cluster and relax time (called $f$ = age/trelax) and a factor $\varpi$, which is related to the total number of observed BSS (NBS). The function is written as $NBS \cong f^3 (\frac{1}{e^{f/\varpi}-1})$. In that paper the results of this equation were corroborated by applying to a couple of OC clusters and, for completing the test, to a set of observed number of BSS of GC listed by MAP. This same catalogue is used in this work but, in order to extend the data, not restricted to only one catalogue this work was also carried out with another BSS catalogue published by LSK. In the case of MAP the adopted number of BSS is displayed within MAP's paper in the table, fourth column under the label NBS,tot. The calculation of the number of theoretical BSS for MAP and LM are done by assuming that the relax time might be the value alike to that one listed as the median relaxation time t(r_h) published by H10 and the median relaxation time from RB in order to avoid bias because of a single source. As GC median age adopted, the value suggested by KC - 12.6 Gyr. Thus, the theoretical number of BSS was computed, cluster by cluster, by means of the above mentioned and for both catalogues: the relation between predicted BSS versus the number of observed BSS was found one to one with an $R^2 \cong 1$

Once the theoretical number of BSS is computed, then the question is to find the relationships between which are linear

**Table 5** List of a complete set of age values from different authors
Table 5 also includes a list of different metallicities

| Cluster name | [Fe/H] H10 | [Fe/H] M&W | [Fe/H] VBLC | t (MWA) | t (MWA) | tmin (S) | t() SW02 | | t(A) BASTI | t(A) BASTI | t(BC) BASTI | t(BC) BASTI | t VBLC | This work (MAP;H10) | This work (MAP;RB) | This work (LSK;H10) | This work (LSK;RB) | age min. Gyr | age Max. Gyr | |
|---|---|---|---|---|---|---|---|---|---|---|---|---|---|---|---|---|---|---|---|---|
| NGC 104 | -0.72 | -0.72 | -0.76 | 11.9 | 13.6 | 11.5 | 10.7 | ±1 | 9.9 | 11.5 | 9.5 | | 11.75 | ++ | ++ | ++ | ++ | | | (**) |
| NGC 362 | -1.26 | -1.26 | -1.3 | 7.2 | 13.2 | 8.5 | 8.7 | ±1.5 | 6.3 | 11.2 | 6.8 | 10.8 | 10.75 | 11.17 | 8.99 | ++ | ++ | 8.99 | 11.17 | |
| NGC 1261 | -1.27 | -1.27 | -1.27 | 8.6 | 14.6 | 9.5 | 8.6 | ±1.1 | 7.7 | 12.2 | 10.5 | 10.6 | 10.75 | 13.49 | ++ | 0.90 | -- | 0.90 | 13.49 | |
| NGC 1851 | -1.18 | -1.18 | -1.18 | 8.8 | 9.9 | 9.5 | 9.2 | ±1.1 | 7.9 | 8.7 | 7.9 | 9.8 | 11 | 13.45 | 12.37 | 11.25 | 10.51 | 10.51 | 13.45 | |
| NGC 1904 | -1.6 | -1.6 | | 10.6 | 12 | 11 | 11.7 | ±1.3 | 10.1 | 11.8 | 12 | | | 10.15 | 8.83 | 11.22 | 9.23 | 8.83 | 11.22 | |
| NGC 2808 | -1.14 | -1.14 | -1.18 | 9.4 | 10 | 10 | 9.3 | ±1.1 | 8.4 | 8.6 | 8 | | 11 | ++ | ++ | 10.48 | ++ | | 10.48 | (**) |
| NGC 3201 | -1.59 | -1.59 | -1.51 | 9 | | 4 | 11.3 | ±1.1 | 7.3 | 8.8 | 4 | | 11.5 | -- | -- | 9.57 | 6.69 | 6.69 | 9.57 | |
| NGC 4147 | -1.8 | -1.8 | -1.78 | | 9.5 | 6 | | | 8.4 | 9 | 4 | | 12.25 | 10.30 | 11.11 | 13.44 | 13.50 | 10.30 | 13.50 | |
| NGC 4372 | | -2.17 | | | | | | | | | | | | -- | -- | 11.30 | 4.66 | 4.66 | 11.30 | |
| NGC 4590 | -2.23 | -2.23 | -2.27 | 8.5 | 11 | 9.5 | 11.2 | ±0.9 | 9 | 12.5 | 9 | | 12 | 2.03 | 1.80 | 13.14 | 11.31 | 2.03 | 13.14 | |
| NGC 4833 | -1.85 | -1.85 | -1.89 | 12.5 | 14 | 11.5 | | | 12.1 | 14.2 | 12 | | 12.5 | 7.88 | 12.01 | ++ | 5.98 | 5.98 | 12.01 | |
| NGC 5024 | -2.1 | -2.1 | -2.06 | | 13 | 11 | | | 11.4 | 13.1 | 10 | | 12.25 | 2.96 | 4.87 | -- | 4.38 | 2.96 | 4.87 | |
| NGC 5634 | -1.88 | -1.88 | | 8.4 | 10.1 | 9 | | | 8.8 | 11.2 | 7.5 | | | ++ | 7.88 | ++ | ++ | | 7.88 | |
| NGC 5694 | -1.98 | -1.98 | | | 12.5 | 13 | | | 11.2 | 13 | 12 | | | 8.27 | 12.06 | 12.34 | 9.97 | 8.27 | 12.34 | (**) |
| NGC 5824 | -1.91 | -1.91 | | 9.2 | 10.4 | 11 | | | 8.8 | 9.5 | 9 | | | 10.16 | 13.06 | ++ | ++ | 10.16 | 13.06 | (**) |
| NGC 5904 | -1.29 | -1.29 | -1.33 | 9.9 | 10.9 | 11 | 10.9 | ±1.1 | 8.7 | 9.7 | 8.5 | 9.6 | 11.5 | 3.98 | ++ | 7.47 | ++ | 3.98 | 7.47 | |
| NGC 5927 | -0.49 | -0.49 | -0.29 | 11.3 | 13.6 | 8 | | | 8.6 | 10.4 | 7 | | 10.75 | 6.07 | 3.79 | 13.37 | 12.86 | 3.79 | 13.37 | |
| NGC 5946 | | -1.29 | | | | | | | | | | | | 3.28 | 2.18 | 1.76 | -- | 1.76 | 3.28 | (**) |
| NGC 5986 | -1.59 | -1.59 | -1.63 | | 12 | 11 | | | 11 | 10.5 | | | 12.25 | 8.91 | 4.19 | 8.58 | 7.24 | 4.19 | 8.91 | (**) |
| NGC 6093 | -1.75 | -1.75 | | | 10 | 10.5 | 12.4 | ±1.1 | 9 | 8 | | | | 11.82 | 11.70 | 12.31 | 11.98 | 11.70 | 12.31 | |
| NGC 6171 | -1.02 | -1.02 | -1.03 | | 9 | 9.5 | 11.7 | ±0.8 | 7.3 | 7.9 | 7.5 | | 12 | -- | -- | 9.60 | 8.05 | 8.05 | 9.60 | |
| NGC 6205 | -1.53 | -1.53 | -1.58 | | 11 | 15.5 | 11.9 | ±1.1 | 9.7 | 10.7 | 10.5 | | 12 | -- | -- | 10.26 | 7.86 | 7.86 | 10.26 | |
| NGC 6218 | | -1.37 | -1.33 | | | | | | | | | | 13 | 9.53 | 7.34 | 11.38 | 9.28 | 7.34 | 11.38 | |
| NGC 6235 | | -1.28 | | | | | | | | | | | | 5.41 | 8.98 | 7.45 | 10.59 | 5.41 | 10.59 | |
| NGC 6266 | -1.18 | -1.18 | | 11 | 12.1 | 13 | | | 10 | 9.5 | | | 12.28 | 3.83 | 10.70 | 8.98 | 3.83 | 12.28 | (**) | |
| NGC 6273 | -1.74 | -1.74 | | 11.1 | 18 | 15 | | | 17 | 17 | | | 12.14 | ++ | ++ | ++ | | 12.14 | | |
| NGC 6284 | -1.26 | -1.26 | | | 11 | 11 | | | 8.9 | 10 | 10 | | 1.34 | -- | -- | -- | 1.34 | | (**) | |
| NGC 6287 | -2.1 | -2.1 | | | 12.5 | 10.5 | | | 10.2 | 11.6 | 9 | | 6.40 | 8.08 | 12.08 | 13.19 | 6.40 | 13.19 | | |
| NGC 6293 | -1.99 | -1.99 | | 7.9 | 8.8 | 10 | | | 8.3 | 8.7 | 8.5 | | 9.92 | | 5.77 | | 5.77 | 9.92 | | |
| NGC 6304 | -0.45 | -0.45 | -0.37 | 12.3 | 14.9 | 10 | | | 10.3 | 12.6 | 9 | | 11.25 | 13.27 | ++ | 12.06 | 11.52 | 11.52 | 13.27 | |
| NGC 6342 | | -0.55 | | | | | | | | | | | | 12.25 | 11.55 | 11.76 | 10.35 | 10.35 | 12.25 | |
| NGC 6356 | -0.4 | -0.4 | | 12 | 18 | 12.5 | | | 10.1 | 15.6 | 10 | | | ++ | 13.44 | 11.32 | 7.76 | 7.76 | 13.44 | |
| NGC 6362 | -0.99 | -0.99 | -1.07 | | 10.5 | 9 | 11 | ±1.3 | 8.5 | 9.5 | 8.5 | | 12.5 | 1.73 | 1.89 | 8.99 | 8.71 | 1.73 | 8.99 | |
| NGC 6388 | | -0.55 | | | | | | | | | | | | ++ | 10.25 | ++ | 11.08 | 10.25 | 11.08 | |
| NGC 6397 | | -2.02 | | | | | | | | | | | 13 | 8.63 | 10.36 | 9.74 | 11.02 | 8.63 | 11.02 | |
| NGC 6441 | | -0.46 | | | | | | | | | | | | ++ | ++ | | | | | |
| NGC 6522 | -1.34 | -1.34 | | 13.9 | 16.1 | 16 | | | 12.2 | 15.4 | 13 | | | | | | | | | |
| NGC 6544 | -1.4 | -1.4 | | 7.1 | 8.5 | 8 | | | 6.7 | 7.5 | 7 | | | 12.46 | 12.47 | 11.22 | 10.99 | 10.99 | 12.47 | |
| NGC 6569 | | -0.76 | | | | | | | | | | | | 8.23 | 3.04 | | | 3.04 | 8.23 | |
| NGC 6584 | -1.5 | -1.5 | -1.5 | | 9 | 10.5 | 11.3 | ±1.4 | 7.9 | 8.5 | 8 | | 11.75 | 7.30 | | 13.16 | | 7.30 | 13.16 | |
| NGC 6624 | -0.44 | -0.44 | -0.42 | | 12 | 10.5 | 10.6 | ±1.4 | | 10 | 8.5 | | 11.25 | 12.63 | 13.06 | 13.31 | 12.44 | 12.44 | 13.31 | |
| NGC 6637 | -0.64 | -0.64 | -0.59 | 12 | 14.5 | 11.5 | 10.6 | ±1.4 | | 10 | 9.5 | | 11 | 12.38 | 12.77 | 13.21 | 11.14 | 11.14 | 13.21 | |
| NGC 6638 | -0.95 | -0.95 | | | 12 | 11.5 | | | 10 | 12.1 | 10 | | | 12.13 | 12.13 | 12.41 | 12.80 | 12.13 | 12.80 | |
| NGC 6642 | -1.26 | -1.26 | | | 10 | 10 | | | | 8.5 | 8.5 | | | 13.51 | 12.09 | 13.51 | 12.22 | 12.09 | 13.51 | |
| NGC 6652 | -0.81 | -0.81 | -0.76 | 11.5 | 11.8 | 8.5 | 11.4 | ±1.0 | 9.1 | 10.8 | 7.5 | | 11.25 | 12.46 | 11.52 | ++ | 13.56 | 11.52 | 13.66 | |
| NGC 6681 | -1.62 | -1.62 | -1.62 | 10.9 | 13.3 | 9.5 | 11.5 | ±1.4 | 10.6 | 13 | 9.5 | | 12.75 | 9.03 | 5.77 | 9.60 | 5.82 | 5.77 | 9.60 | |
| NGC 6712 | -1.02 | | -1.44 | 10.2 | 14.7 | 9.5 | 10.4 | ±1.4 | 8.7 | 12.1 | 9.5 | | | | | | | | | |
| NGC 6717 | | -1.26 | -1.26 | | | | | | | | | | 12.5 | 13.36 | 13.19 | 12.94 | 13.19 | 12.94 | 13.36 | |
| NGC 6723 | -1.1 | -1.1 | -1.1 | | 10 | 10.5 | 11.6 | ±1.3 | 8.3 | 9.5 | 8 | | 12.5 | 1.90 | 1.85 | 8.93 | 7.86 | 1.85 | 8.93 | |
| NGC 6838 | -0.78 | -0.78 | -0.82 | | 12 | 11.5 | 10.2 | ±1.4 | | 10 | 10 | | 11 | 12.56 | 13.15 | 12.32 | 12.92 | 12.32 | 13.15 | |
| NGC 6864 | -1.29 | -1.29 | | | 9 | 9.5 | | | | 8 | 7.5 | | | 12.41 | 6.98 | ++ | 10.66 | 6.98 | 12.41 | |
| NGC 6934 | -1.47 | -1.47 | -1.56 | 8.7 | 10.2 | 9.5 | 9.6 | ±1.5 | 7.8 | 8.7 | 9 | | 11.75 | 4.42 | 2.88 | 12.15 | 10.12 | 2.88 | 12.15 | |
| NGC 6981 | -1.42 | -1.42 | -1.48 | | 9.5 | 9.5 | | | 8.4 | 8.5 | 7.9 | 8.5 | 11.5 | 4.35 | 3.88 | 7.28 | 11.18 | 3.88 | 11.18 | |
| NGC 7078 | -2.37 | | -2.33 | 8.6 | 9.2 | 11 | 11.7 | ±0.8 | | 8 | 9 | | 12.75 | ++ | 12.99 | 1.43 | -- | 1.43 | 12.99 | (**) |
| NGC 7089 | -1.65 | -1.65 | -1.66 | 11.8 | 13.7 | 12.5 | | | 10.9 | 11.6 | 11 | | 11.75 | -- | -- | -- | -- | | | |
| NGC 7099 | -2.27 | -2.27 | -2.33 | 13.7 | 15.4 | 13 | 11.9 | ±1.4 | 12.8 | 15 | 14 | | 13 | 9.75 | 8.96 | 9.54 | 7.68 | 7.68 | 9.75 | (**) |
| **Median** | **-1.37** | **-1.29** | **-1.33** | *10.4* | *12* | *10.5* | *11.1* | | *8.8* | *10.2* | *9* | *9.8* | *11.75* | **9.84** | **8.99** | **11.24** | **10.55** | **7.32** | **12.15** | |

(**) deserve separate analysis. -- means computed age value below cero '++ means computed age value over 13.6 Gyr Values marked in slight dark generating the major doubt according to the work performed by LM. The mentioned linear relation is written as ($\varpi = A*f + B$). Looking for the best fit; thus a new value of $f$ is derived, named $f_T$, which allows deriving the GC age by means the following equation: age= $f_T$ * trlx.

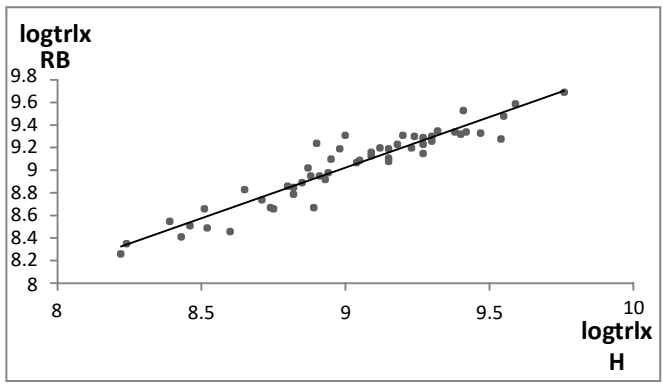

Fig.2. The relation between the two quoted values for trlx is depicted. The straight line represents that the correlation is lineal with a goodness of fit of 90%.

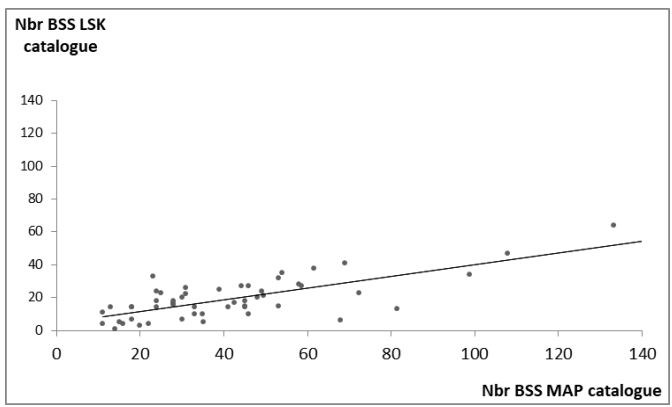

Fig.3. This figures represents the difference between the number of BSS for each cluster and listed in the two catalogues used in this work (MAP & LSK). It has been drawn the two axis with the same scale for enhancing the mentioned difference.

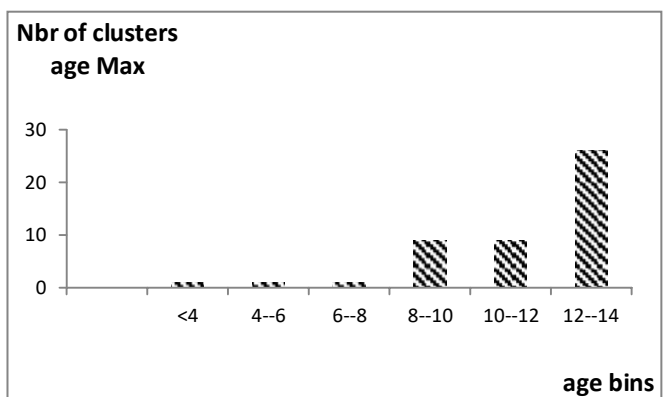

Fig.4. This figure represents the number of cluster per each age interval.

As the age of GC has to be smaller (compatible) than the age for the Universe of ~ 13.8 Gyr, consequently the GCs have not to be older than 13.6 Gyr; that is translated to put aside clusters whose computed ages are older than 13.6 Gyr. Newly, the process continues, it is repeated, with the rest of the values. At each step, a new equation is solved followed by another $f_T$ is obtained yielding to a new value for age. In the same way is done if the value of calculated age is negative. All detailed values are displayed in Table 1, Table 2, Table 3 and Table 4.

In order to show that the present reached value match with the common accepted ones Figure 6 displays a histogram reflecting the number of clusters in each interval. More than 50% (55%) of the cluster of this work are centred in an age interval from 12 to 14 Gyr.

All results are displayed in table 5 where it has been included: metallicity from H10, from M&W and from VBLC. As a first conclusion: it is confirmed that this method for age determination is not depending on the metallicity, although the present values are compatible with those obtained by Forbes et al (2015). In order to compare the results of the present work with those obtained by other authors, it has been included all set of ages listed by M&S and by VBLC. As a second conclusion: ages determinate by this work are comparable and compatibles with those published by M&S and by VBLC; however the present method is much simpler.

Finally Table 5 exhibits two last columns which correspond to a proposal: range of minimum and maximum values of ages for the set of GC clusters considered in the present study. The values listed in this last column (age Maximum) of table 5 have been compared with those published by VandenBerg et al (2013); the average of the difference for those common clusters is ~0.22 Gyr.

Nevertheless there are some unruly clusters: NGC 5946, NGC 5986, NGC 6284, NGC6362, NGC6723 and NGC 7078) whose calculated ages are when less, very strange. The main reason there is to find it in the likely wrong number of the observed BSS.

Those clusters that present ages older than 13.6 Gyr (marked with ++ in Table 5), actually only present discrepancies in the partial age calculations depending on the reference authors. (see section 2: MAP;H10, MAP;RB, LSK;H10 and LSK;RB).

The most surprising cluster is NGC 104, whose age exceeds 13.6 Gyr (age in between 19.04 and 20.30Gyr), which would have a very exotic explanation: the existence of an intermediate black hole in the center of the cluster as it is suggested by Kizilton et al 2017. Recently Ke Qin et al 2023 simulations showed that black hole main-sequence star (BH-MS) binaries with an initial orbital period less than the bifurcation period can evolve into ultra-compact X-ray binaries (UCXBs) that can be detected by LISA. On the other hand, if that age were true, it would call into question the expansion age for a flat Universe. This would call into question the case for a dark energy–dominated Universe.

***Acknowledgements:*** F. Llorente de Andrés acknowledges support from the Faculty of the European Space Astronomy Centre (ESAC), thanks A. Parras (CAB) and M. Sánchez (POLATEC) for their helpful technical collaboration. The author also recognizes the fruitful discussions with Dr. J.A. Caballero who gave many suggestions and recommendations.

**Table 1** Age of clusters. BSS are quoted from MAP catalogue and values of trlx from H10.

It is split into: Table 1a, Table 1b, Table 1c, Table 1d and Table 1e. There are some ages which are negatives but they are listed for further analysis.

Table 1a

$f_{1T} < 6 // \varpi_1 \geq 4$   Transform $f_{1T} = 7.08 - \varpi_1/1.98$ fit $R^2 = 0.52$

| Cluster name | logtrlx H10 | age | $f_1$ | $\varpi_1$ | NBS cal1 | NBS obs | $f_{1T}$ TOT | $age_{1T}$ |
|---|---|---|---|---|---|---|---|---|
| IC 4499 | 9.73 | 12.6 | 2.35 | 7 | 32 | 33 | 3.54 | 19.04 |
| NGC 104 | 9.55 | 12.6 | 3.55 | 5.8 | 53 | 54 | 4.15 | 14.73 |
| NGC 5024 | 9.76 | 12.6 | 2.19 | 13 | 57 | 58 | 0.51 | 2.96 |
| NGC 5634 | 9.54 | 12.6 | 3.63 | 5 | 45 | 46 | 4.55 | 15.79 |
| NGC 5824 | 9.47 | 12.6 | 4.27 | 7.2 | 96 | 98.7 | 3.44 | 10.16 |
| NGC 6273 | 9.38 | 12.6 | 5.25 | 4 | 53 | 53.1 | 5.06 | 12.14 |
| **Median** | | | | **6.4** | **53** | **53.6** | **3.85** | **13.43** |

Table 1b

Transform $f_{1T} = \varpi_1/0.068 - 37.143$ fit $R^2 = 0.79$

| Cluster name | logtrlx H10 | age | $f_1$ | $\varpi_1$ | NBS cal1 | NBS obs | $f_{1T}$ TOT | $age_{1T}$ |
|---|---|---|---|---|---|---|---|---|
| NGC 1904 | 8.95 | 12.6 | 14.14 | 3.3 | 39 | 39 | 11.39 | 10.15 |
| NGC 4147 | 8.74 | 12.6 | 22.93 | 3.8 | 29 | 28 | 18.74 | 10.30 |
| NGC 4590 | 9.27 | 12.6 | 6.77 | 2.6 | 25 | 24 | 1.09 | 2.03 |
| NGC 5927 | 8.94 | 12.6 | 14.47 | 3 | 25 | 23 | 6.97 | 6.07 |
| NGC 5946 | 8.91 | 12.6 | 15.50 | 2.8 | 15 | 14 | 4.03 | 3.28 |
| NGC 6093 | 8.8 | 12.6 | 19.97 | 3.8 | 42 | 44.3 | 18.74 | 11.82 |
| NGC 6171 | 9 | 12.6 | 12.60 | 3.1 | 35 | 33 | 8.45 | 8.45 |
| NGC 6218 | 8.87 | 12.6 | 17.00 | 3.4 | 33 | 31 | 12.86 | 9.53 |
| NGC 6235 | 8.89 | 12.6 | 16.23 | 3 | 19 | 18 | 6.97 | 5.41 |
| NGC 6266 | 8.98 | 12.6 | 13.19 | 3.4 | 48 | 49.1 | 12.86 | 12.28 |
| NGC 6284 | 9.09 | 12.6 | 10.24 | 2.6 | 21 | 22 | 1.09 | 1.34 |
| NGC 6287 | 8.75 | 12.6 | 22.41 | 3.3 | 13 | 13 | 11.39 | 6.40 |
| NGC 6293 | 8.94 | 12.6 | 14.47 | 3.3 | 38 | 35.1 | 11.39 | 9.92 |
| NGC 6304 | 8.85 | 12.6 | 17.80 | 3.8 | 53 | 53 | 18.74 | 13.27 |
| NGC 6342 | 8.51 | 12.6 | 38.94 | 5.1 | 29 | 30.1 | 37.86 | 12.25 |
| NGC 6362 | 9.2 | 12.6 | 7.95 | 2.6 | 25 | 25 | 1.09 | 1.73 |
| NGC 6397 | 8.6 | 12.6 | 31.65 | 4 | 12 | 11 | 21.68 | 8.63 |
| NGC 6544 | 8.24 | 12.6 | 72.51 | 7.4 | 21 | 20 | 71.68 | 12.46 |
| NGC 6584 | 9.02 | 12.6 | 12.03 | 3 | 32 | 31 | 6.97 | 7.30 |
| NGC 6624 | 8.71 | 12.6 | 24.57 | 4.2 | 43 | 45 | 24.62 | 12.63 |
| NGC 6637 | 8.82 | 12.6 | 19.07 | 3.8 | 46 | 48 | 18.74 | 12.38 |
| NGC 6642 | 8.52 | 12.6 | 38.05 | 5.3 | 42 | 41 | 40.80 | 13.51 |
| NGC 6681 | 8.65 | 12.6 | 28.21 | 3.9 | 16 | 16 | 20.21 | 9.03 |
| NGC 6717 | 8.22 | 12.6 | 75.92 | 8 | 33 | 33 | 80.50 | 13.36 |
| NGC 6723 | 9.24 | 12.6 | 7.25 | 2.6 | 25 | 24 | 1.09 | 1.90 |
| NGC 6838 | 8.43 | 12.6 | 46.81 | 5.7 | 28 | 28 | 46.68 | 12.56 |
| NGC 6934 | 9.04 | 12.6 | 11.49 | 2.8 | 25 | 24 | 4.03 | 4.42 |
| NGC 6981 | 9.23 | 12.6 | 7.42 | 2.7 | 28 | 28 | 2.56 | 4.35 |
| NGC 7099 | 8.88 | 12.6 | 16.61 | 3.4 | 35 | 35 | 12.86 | 9.75 |
| **Median** | | | | **3.4** | **29** | **28** | **12.86** | **9.53** |

Table 1c

Transform $f_{1T} = \varpi_1/0.069 - 41.868$ fit $R^2 = 0.93$

| Cluster name | logtrlx H10 | age | $f_1$ | $\varpi_1$ | NBS cal1 | NBS obs | $f_{1T}$ TOT | $age_{1T}$ |
|---|---|---|---|---|---|---|---|---|
| NGC 362 | 8.93 | 12.6 | 14.80 | 3.8 | 67 | 69 | 13.12 | 11.17 |
| NGC 1261 | 9.12 | 12.6 | 9.56 | 3.6 | 66 | 68 | 10.23 | 13.49 |
| NGC 1851 | 8.82 | 12.6 | 19.07 | 4.3 | 83 | 81.4 | 20.36 | 13.45 |
| NGC 4833 | 9.42 | 12.6 | 4.79 | 3.1 | 30 | 30 | 2.99 | 7.88 |
| NGC 5694 | 9.27 | 12.6 | 6.77 | 3.2 | 43 | 42.5 | 4.44 | 8.27 |
| NGC 5904 | 9.41 | 12.6 | 4.90 | 3 | 29 | 28 | 1.55 | 3.98 |
| NGC 5986 | 9.18 | 12.6 | 8.32 | 3.3 | 50 | 49.6 | 5.89 | 8.91 |
| NGC 6569 | 9.05 | 12.6 | 11.23 | 3.4 | 54 | 52.7 | 7.34 | 8.23 |
| NGC 6638 | 8.46 | 12.6 | 43.69 | 5.8 | 45 | 45 | 42.07 | 12.13 |
| NGC 6652 | 8.39 | 12.6 | 51.33 | 6.4 | 44 | 45 | 50.75 | 12.46 |
| NGC 6864 | 9.15 | 12.6 | 8.92 | 3.5 | 60 | 58.6 | 8.78 | 12.41 |
| **Median** | | | | **3.5** | **50** | **49.6** | **8.78** | **11.17** |

Table 1d

| Cluster name | logtrlx H10 | age | $f_1$ | $\varpi_1$ | NBS cal1 | NBS obs | $f_{1T}$ TOT | $age_{1T}$ |
|---|---|---|---|---|---|---|---|---|
| NGC 2808 | 9.15 | 12.6 | 8.92 | 4.4 | 108 | 107.8 | 27.56 | 38.93 |
| NGC 6273 | 9.38 | 12.6 | 5.25 | 4 | 53 | 53.1 | 21.68 | 52.01 |
| NGC 6356 | 9.3 | 12.6 | 6.31 | 4.2 | 72 | 72.4 | 24.62 | 49.13 |
| NGC 6441 | 9.09 | 12.6 | 10.24 | 4.8 | 144 | 143 | 33.45 | 41.15 |
| NGC 6229 | 9.15 | 12.6 | 8.92 | 3.9 | 80 | 79.7 | 14.57 | 20.58 |
| NGC 6388 | 8.9 | 12.6 | 15.86 | 4.6 | 131 | 133.2 | 24.7 | 19.62 |
| NGC 7078 | 9.32 | 12.6 | 6.03 | 3.4 | 45 | 46 | 7.336 | 15.33 |
| **Median** | | | | **4.2** | **80** | **79.7** | **24.62** | **38.93** |

Table 1e

| Cluster name | logtrlx H10 | age | $f_1$ | $\varpi_1$ | NBS cal1 | NBS obs | $f_{1T}$ TOT | $age_{1T}$ |
|---|---|---|---|---|---|---|---|---|
| NGC 3201 | 9.27 | 12.6 | 6.77 | 2.3 | 17 | 18 | -3.32 | -6.18 |
| NGC 4372 | 9.59 | 12.6 | 3.24 | 2.3 | 11 | 11 | -3.32 | -12.9 |
| NGC 6205 | 9.3 | 12.6 | 6.31 | 2.3 | 17 | 18 | -3.32 | -6.62 |
| NGC 7089 | 9.4 | 12.6 | 5.02 | 2.2 | 14 | 15 | -4.79 | -12 |
| **Median** | | | | **2.3** | **16** | **17** | **-3.32** | **-9.33** |

**Table 2** Age of clusters. BSS are quoted from MAP catalogue and values of trlx from RB.
.It is split into: Table 2a, Table 2b, Table 2c, Table 2d, Table 2e
There are some ages which are negatives but they are listed for further analysis.

Table 2a

| Cluster name | logtrlx RB | age | $f_1$ | $\varpi_1$ | NBS cal1 | NBS obs TOT | $f_{1T}$ | age$_{1T}$ |
|---|---|---|---|---|---|---|---|---|
| NGC 104  | 9.48 | 12.6 | 4.17 | 4.9  | 54  | 54    | 6.43 | 19.41 |
| NGC 5024 | 9.69 | 12.6 | 2.57 | 10   | 58  | 58    | 0.99 | 4.87  |
| NGC 5824 | 9.33 | 12.6 | 5.89 | 5.2  | 97  | 98.7  | 6.11 | 13.06 |
| NGC 6356 | 9.26 | 12.6 | 6.92 | 4    | 71  | 72.4  | 7.39 | 13.44 |
| NGC 6388 | 9.24 | 12.6 | 7.25 | 5.4  | 135 | 133.2 | 5.90 | 10.25 |
| Mesdian  |      |      |      | **5.2** | **71** | **72.4** | **6.11** | **13.06** |

Table 2b

Transform $f_{1T}= \varpi_1/0.0657 - 38.6484$  fit $R^2 = 0.79$

| Cluster name | logtrlx RB | age | $f_1$ | $\varpi_1$ | NBS cal1 | NBS obs TOT | $f_{1T}$ | age$_{1T}$ |
|---|---|---|---|---|---|---|---|---|
| NGC 1904 | 9.1  | 12.6 | 10.01 | 3   | 37 | 39   | 7.01  | 8.83  |
| NGC 4147 | 8.67 | 12.6 | 26.94 | 4.1 | 27 | 28   | 23.76 | 11.11 |
| NGC 4590 | 9.29 | 12.6 | 6.46  | 2.6 | 25 | 24   | 0.93  | 1.80  |
| NGC 4833 | 9.34 | 12.6 | 5.76  | 2.9 | 30 | 30   | 5.49  | 12.01 |
| NGC 5694 | 9.15 | 12.6 | 8.92  | 3.1 | 42 | 42.5 | 8.54  | 12.06 |
| NGC 5927 | 8.98 | 12.6 | 13.19 | 2.8 | 21 | 23   | 3.97  | 3.79  |
| NGC 5946 | 8.95 | 12.6 | 14.14 | 2.7 | 15 | 14   | 2.45  | 2.18  |
| NGC 6093 | 8.86 | 12.6 | 17.39 | 3.6 | 42 | 44.3 | 16.15 | 11.70 |
| NGC 6218 | 9.02 | 12.6 | 12.03 | 3   | 32 | 31   | 7.01  | 7.34  |
| NGC 6235 | 8.67 | 12.6 | 26.94 | 3.8 | 16 | 18   | 19.19 | 8.98  |
| NGC 6287 | 8.66 | 12.6 | 27.57 | 3.7 | 12 | 13   | 17.67 | 8.08  |
| NGC 6304 | 8.89 | 12.6 | 16.23 | 3.7 | 54 | 53   | 17.67 | 13.71 |
| NGC 6342 | 8.66 | 12.6 | 27.57 | 4.2 | 30 | 30.1 | 25.28 | 11.55 |
| NGC 6362 | 9.31 | 12.6 | 6.17  | 2.6 | 24 | 25   | 0.93  | 1.89  |
| NGC 6397 | 8.46 | 12.6 | 43.69 | 4.9 | 11 | 11   | 35.93 | 10.36 |
| NGC 6544 | 8.35 | 12.6 | 56.28 | 6.2 | 20 | 20   | 55.72 | 12.47 |
| NGC 6624 | 8.74 | 12.6 | 22.93 | 4.1 | 45 | 45   | 23.76 | 13.06 |
| NGC 6637 | 8.79 | 12.6 | 20.43 | 3.9 | 45 | 48   | 20.71 | 12.77 |
| NGC 6681 | 8.83 | 12.6 | 18.64 | 3.1 | 16 | 16   | 8.54  | 5.77  |
| NGC 6723 | 9.3  | 12.6 | 6.31  | 2.6 | 24 | 24   | 0.93  | 1.85  |
| NGC 6838 | 8.41 | 12.6 | 49.02 | 5.9 | 29 | 28   | 51.15 | 13.15 |
| NGC 6934 | 9.07 | 12.6 | 10.72 | 2.7 | 24 | 24   | 2.45  | 2.88  |
| NGC 6981 | 9.2  | 12.6 | 7.95  | 2.7 | 28 | 28   | 2.45  | 3.88  |
| NGC 7099 | 8.95 | 12.6 | 14.14 | 3.2 | 34 | 35   | 10.06 | 8.96  |
| Median   |      |      |       | **3.15** | **28** | **28** | **9.30** | **8.97** |

Table 2c

Transform $f_{1T}= \varpi_1/0.06 - 52.53$  fit $R^2 = 0.87$

| Cluster name | logtrlx RB | age | $f_1$ | $\varpi_1$ | NBS cal1 | NBS obs TOT | $f_{1T}$ | age$_{1T}$ |
|---|---|---|---|---|---|---|---|---|
| NGC 362  | 8.92 | 12.6 | 15.15 | 3.8 | 66 | 69   | 10.80 | 8.99  |
| NGC 1851 | 8.85 | 12.6 | 17.80 | 4.2 | 83 | 81.4 | 17.47 | 12.37 |
| NGC 5634 | 9.28 | 12.6 | 6.61  | 3.4 | 48 | 46   | 4.14  | 7.88  |
| NGC 5986 | 9.23 | 12.6 | 7.42  | 3.3 | 48 | 49.6 | 2.47  | 4.19  |
| NGC 6266 | 9.19 | 12.6 | 8.14  | 3.3 | 50 | 49.1 | 2.47  | 3.83  |
| NGC 6569 | 9.09 | 12.6 | 10.24 | 3.3 | 50 | 52.7 | 2.47  | 3.04  |
| NGC 6638 | 8.51 | 12.6 | 38.94 | 5.4 | 44 | 45   | 37.47 | 12.13 |
| NGC 6642 | 8.49 | 12.6 | 40.77 | 5.5 | 41 | 41   | 39.14 | 12.09 |
| NGC 6652 | 8.55 | 12.6 | 35.51 | 5.1 | 42 | 45   | 32.47 | 11.52 |
| NGC 6717 | 8.26 | 12.6 | 69.24 | 7.5 | 32 | 33   | 72.47 | 13.19 |
| NGC 6864 | 9.08 | 12.6 | 10.48 | 3.5 | 61 | 58.6 | 5.80  | 6.98  |
| NGC 7078 | 9.35 | 12.6 | 5.63  | 3.5 | 45 | 46   | 5.80  | 12.99 |
| Median   |      |      |       | **3.65** | **48** | **48** | **8.30** | **10.25** |

Table 2d

| Cluster name | logtrlx RB | age | $f_1$ | $\varpi_1$ | NBS cal1 | NBS obs TOT | $f_{1T}$ | age$_{1T}$ |
|---|---|---|---|---|---|---|---|---|
| NGC 1261 | 9.2  | 12.6 | 7.95 | 3.7 | 66  | 68    | 9.14  | 14.48 |
| NGC 2808 | 9.11 | 12.6 | 9.78 | 4.3 | 107 | 107.8 | 19.14 | 24.65 |
| NGC 5904 | 9.53 | 12.6 | 3.72 | 3.4 | 26  | 28    | 4.14  | 14.02 |
| NGC 6229 | 9.19 | 12.6 | 8.14 | 4   | 81  | 79.7  | 14.14 | 21.90 |
| NGC 6273 | 9.34 | 12.6 | 5.76 | 3.8 | 54  | 53.1  | 10.80 | 23.64 |
| NGC 6441 | 9.13 | 12.6 | 9.34 | 4.9 | 142 | 143   | 29.14 | 39.30 |
| Median   |      |      |      | **3.9** | **74** | **73.85** | **12.47** | **22.77** |

Table 2e

| Cluster name | logtrlx RB | age | $f_1$ | $\varpi_1$ | NBS cal1 | NBS obs TOT | $f_{1T}$ | age$_{1T}$ |
|---|---|---|---|---|---|---|---|---|
| NGC 3201 | 9.23 | 12.6 | 7.42 | 2.3 | 17 | 18 | -3.64 | -6.18   |
| NGC 4372 | 9.59 | 12.6 | 3.24 | 2.3 | 11 | 11 | -3.64 | -14.16  |
| NGC 6205 | 9.3  | 12.6 | 6.31 | 2.3 | 17 | 18 | -3.64 | -7.26   |
| NGC 6284 | 9.16 | 12.6 | 8.72 | 2.5 | 21 | 22 | -0.60 | -0.86   |
| NGC 7089 | 9.32 | 12.6 | 6.03 | 2.2 | 15 | 15 | -5.16 | -10.79  |
| NGC 6171 | 9.31 | 12.6 | 6.17 | 3   | 34 | 33 | -2.53 | -5.17   |
| median   |      |      |      | **2.3** | **17.1** | **18** | **-3.64** | **-6.72** |

**Table 3** Age of clusters. BSS are quoted from LSK catalogue and values of trlx from H10.
.It is split into: Table 3a, Table 3b, Table 3c, Table 3d, Table 3e. There are some ages which are negatives but they are listed for further analysis.

Table 3a

f2T <9// ϖ2 ≥ 3    Transform f2T= 16.961 -ϖ2/0.3354  fit R² = 0.48

| Cluster | logtrlx H10 | age | f2T | ϖ2 | NBS | NBS obs | f2T | age2T |
|---|---|---|---|---|---|---|---|---|
| NGC 104 | 9.55 | 12.6 | 3.55 | 4.3 | 35 | 35 | 4.14 | 14.69 |
| NGC 2808 | 9.15 | 12.6 | 8.92 | 3.2 | 47 | 47 | 7.42 | 10.48 |
| NGC 5634 | 9.54 | 12.6 | 3.63 | 3.6 | 28 | 27 | 6.23 | 21.59 |
| NGC 5824 | 9.47 | 12.6 | 4.27 | 3.6 | 34 | 34 | 6.23 | 18.38 |
| NGC 6229 | 9.15 | 12.6 | 8.92 | 3 | 38 | 38 | 8.02 | 11.32 |
| NGC 6273 | 9.38 | 12.6 | 5.25 | 3.1 | 33 | 32 | 7.72 | 18.51 |
| Median | | | | **3.4** | **35** | **34.5** | **6.82** | **16.53** |

Table 3b

Transform f2T= ϖ2/0.0673- 27.548 fit R² = 0.79

| Cluster name | logtrlx H10 | age | f2T | ϖ2 | NBS cal2 | NBS obs TOT | f2T | age2T |
|---|---|---|---|---|---|---|---|---|
| NGC 1261 | 9.12 | 12.6 | 9.56 | 1.9 | 6 | 6 | 0.68 | 0.90 |
| NGC 1851 | 8.82 | 12.6 | 19.07 | 3 | 12 | 13 | 17.03 | 11.25 |
| NGC 3201 | 9.27 | 12.6 | 6.77 | 2.2 | 15 | 14 | 5.14 | 9.57 |
| NGC 4147 | 8.74 | 12.6 | 22.93 | 3.5 | 17 | 16 | 24.46 | 13.44 |
| NGC 5694 | 9.27 | 12.6 | 6.77 | 2.3 | 17 | 17 | 6.63 | 12.34 |
| NGC 5946 | 8.91 | 12.6 | 15.50 | 2 | 2 | 1 | 2.17 | 1.76 |
| NGC 6171 | 9 | 12.6 | 12.60 | 2.5 | 13 | 14 | 9.60 | 9.60 |
| NGC 6205 | 9.3 | 12.6 | 6.31 | 2.2 | 15 | 14 | 5.14 | 10.26 |
| NGC 6235 | 8.89 | 12.6 | 16.23 | 2.5 | 6 | 7 | 9.60 | 7.45 |
| NGC 6287 | 8.75 | 12.6 | 22.41 | 3.3 | 13 | 14 | 21.49 | 12.08 |
| NGC 6293 | 8.94 | 12.6 | 14.47 | 2.3 | 6 | 5 | 6.63 | 5.77 |
| NGC 6304 | 8.85 | 12.6 | 17.80 | 3 | 15 | 15 | 17.03 | 12.06 |
| NGC 6342 | 8.51 | 12.6 | 38.94 | 4.3 | 7 | 7 | 36.35 | 11.76 |
| NGC 6397 | 8.6 | 12.6 | 31.65 | 3.5 | 4 | 4 | 24.46 | 9.74 |
| NGC 6544 | 8.24 | 12.6 | 72.51 | 6.2 | 3 | 3 | 64.58 | 11.22 |
| NGC 6584 | 9.02 | 12.6 | 12.03 | 2.7 | 20 | 22 | 12.57 | 13.16 |
| NGC 6624 | 8.71 | 12.6 | 24.57 | 3.6 | 16 | 15 | 25.94 | 13.31 |
| NGC 6637 | 8.82 | 12.6 | 19.07 | 3.2 | 18 | 20 | 20.00 | 13.21 |
| NGC 6642 | 8.52 | 12.6 | 38.05 | 4.6 | 14 | 14 | 40.80 | 13.51 |
| NGC 6652 | 8.39 | 12.6 | 51.33 | 5.6 | 14 | 14 | 55.66 | 13.66 |
| NGC 6681 | 8.65 | 12.6 | 28.21 | 3.3 | 4 | 4 | 21.49 | 9.60 |
| NGC 6717 | 8.22 | 12.6 | 75.92 | 7.1 | 10 | 10 | 77.95 | 12.94 |
| NGC 6723 | 9.24 | 12.6 | 7.25 | 2.2 | 15 | 14 | 5.14 | 8.93 |
| NGC 6934 | 9.04 | 12.6 | 11.49 | 2.6 | 18 | 18 | 11.08 | 12.15 |
| NGC 7078 | 9.32 | 12.6 | 6.03 | 1.9 | 10 | 10 | 0.68 | 1.43 |
| NGC 7099 | 8.88 | 12.6 | 16.61 | 2.7 | 10 | 10 | 12.57 | 9.54 |
| Median | | | | **2.85** | **13** | **14** | **14.80** | **11.24** |

Table 3c

Transform f2T= ϖ2/0.0723- 28.907 fit R² = 0.96

| Cluster name | logtrlx H10 | age | f2T | ϖ2 | NBS cal2 | NBS obs TOT | f2T | age2T |
|---|---|---|---|---|---|---|---|---|
| NGC 1904 | 8.95 | 12.6 | 14.14 | 3 | 26 | 25 | 12.59 | 11.22 |
| NGC 4372 | 9.59 | 12.6 | 3.24 | 2.3 | 11 | 11 | 2.90 | 11.30 |
| NGC 4590 | 9.27 | 12.6 | 6.77 | 2.6 | 25 | 24 | 7.05 | 13.14 |
| NGC 5904 | 9.41 | 12.6 | 4.90 | 2.3 | 16 | 16 | 2.90 | 7.47 |
| NGC 5927 | 8.94 | 12.6 | 14.47 | 3.2 | 33 | 33 | 15.35 | 13.37 |
| NGC 5986 | 9.18 | 12.6 | 8.32 | 2.5 | 21 | 21 | 5.67 | 8.58 |
| NGC 6093 | 8.8 | 12.6 | 19.97 | 3.5 | 27 | 27 | 19.50 | 12.31 |
| NGC 6218 | 8.87 | 12.6 | 17.00 | 3.2 | 24 | 26 | 15.35 | 11.38 |
| NGC 6266 | 8.98 | 12.6 | 13.19 | 2.9 | 25 | 24 | 11.20 | 10.70 |
| NGC 6356 | 9.3 | 12.6 | 6.31 | 2.5 | 22 | 23 | 5.67 | 11.32 |
| NGC 6362 | 9.2 | 12.6 | 7.95 | 2.5 | 22 | 23 | 5.67 | 8.99 |
| NGC 6638 | 8.46 | 12.6 | 43.69 | 5.2 | 19 | 18 | 43.02 | 12.41 |
| NGC 6838 | 8.43 | 12.6 | 46.81 | 5.4 | 18 | 17 | 45.78 | 12.32 |
| NGC 6981 | 9.23 | 12.6 | 7.42 | 2.4 | 19 | 18 | 4.29 | 7.28 |
| Median | | | | **2.75** | **22** | **23** | **9.13** | **11.31** |

Table 3d

| Cluster name | logtrlx H10 | age | f2T | ϖ2 | NBS cal2 | NBS obs TOT | f2T | age2T |
|---|---|---|---|---|---|---|---|---|
| NGC 362 | 8.93 | 12.6 | 14.80 | 3.4 | 42.25 | 41 | 18.12 | 15.42 |
| NGC 4833 | 9.42 | 12.6 | 4.79 | 2.6 | 20.69 | 20 | 7.05 | 18.55 |
| NGC 6388 | 8.9 | 12.6 | 15.86 | 3.8 | 62.37 | 64 | 23.65 | 18.79 |
| NGC 6402 | 9.39 | 12.6 | 5.13 | 2.5 | 19.91 | 19 | 5.67 | 13.92 |
| NGC 6712 | 8.95 | 12.6 | 14.14 | 3.3 | 39.50 | 39 | 16.74 | 14.92 |
| NGC 6864 | 9.15 | 12.6 | 8.92 | 2.8 | 30.61 | 27 | 9.82 | 13.87 |
| Median | | | | **3.05** | **35.06** | **33** | **13.28** | **15.17** |

Table 3e

| Cluster name | logtrlx H10 | age | f2T | ϖ2 | NBS cal2 | NBS obs TOT | f2T | age2T |
|---|---|---|---|---|---|---|---|---|
| NGC 5024 | 9.76 | 12.6 | 2.19 | 6.7 | 27 | 28 | -3.02 | -17.35 |
| NGC 6284 | 9.09 | 12.6 | 10.24 | 1.8 | 4 | 4 | -0.80 | -0.99 |
| NGC 7089 | 9.4 | 12.6 | 5.02 | 1.5 | 5 | 5 | -5.26 | -13.21 |
| Median | | | | **1.8** | **4.62** | **5** | **-3.02** | **-13.21** |

**Table 4** Age of clusters. BSS are quoted from LSK catalogue and values of trlx from RB.
It is split into: Table 4a, Table 4b, Table 4c, Table 4d, Table 4e.
There are some ages which are negatives but they are listed for further analysis.

Table 4a

Transform $f_{2T}=19.401 -\varpi_2/0.2918$ fit $R^2 = 0.53$

| Cluster | logtrlx RB | age | $f_{2T}$ | $\varpi_2$ | NBS | NBS obs | $f_{2T}$ | age2T |
|---|---|---|---|---|---|---|---|---|
| NGC 104 | 9.48 | 12.6 | 4.17 | 3.7 | 35 | 35 | 6.72 | 20.30 |
| NGC 5024 | 9.69 | 12.6 | 2.57 | 5.4 | 28 | 28 | 0.90 | 4.38 |
| NGC 6388 | 9.24 | 12.6 | 7.25 | 3.8 | 66 | 64 | 6.38 | 11.08 |
| Median | | | | 3.8 | 35 | 35 | 6.38 | 11.08 |

Table 4b

Transform $f_{2T}= \varpi_2/0.0642- 30.327$ fit $R^2 = 0.86$

| Cluster name | logtrlx RB | h age2 | $f_{2T}$ | $\varpi_2$ | NBS cal2 | N BSS | $f_{2T}$ | age2T |
|---|---|---|---|---|---|---|---|---|
| NGC 1851 | 8.85 | 12.6 | 17.80 | 2.9 | 12 | 13 | 14.84 | 10.51 |
| NGC 3201 | 9.23 | 12.6 | 7.42 | 2.2 | 15 | 14 | 3.94 | 6.69 |
| NGC 4147 | 8.67 | 12.6 | 26.94 | 3.8 | 16 | 16 | 28.86 | 13.50 |
| NGC 5694 | 9.15 | 12.6 | 8.92 | 2.4 | 18 | 17 | 7.06 | 9.97 |
| NGC 6171 | 9.31 | 12.6 | 6.17 | 2.2 | 15 | 14 | 3.94 | 8.05 |
| NGC 6205 | 9.3 | 12.6 | 6.31 | 2.2 | 15 | 14 | 3.94 | 7.86 |
| NGC 6235 | 8.67 | 12.6 | 26.94 | 3.4 | 7 | 7 | 22.63 | 10.59 |
| NGC 6287 | 8.66 | 12.6 | 27.57 | 3.8 | 15 | 14 | 28.86 | 13.19 |
| NGC 6304 | 8.89 | 12.6 | 16.23 | 2.9 | 16 | 15 | 14.84 | 11.52 |
| NGC 6342 | 8.66 | 12.6 | 27.57 | 3.4 | 6 | 7 | 22.63 | 10.35 |
| NGC 6397 | 8.46 | 12.6 | 43.69 | 4.4 | 4 | 4 | 38.21 | 11.02 |
| NGC 6544 | 8.35 | 12.6 | 56.28 | 5.1 | 3 | 3 | 49.11 | 10.99 |
| NGC 6624 | 8.74 | 12.6 | 22.93 | 3.4 | 14 | 15 | 22.63 | 12.44 |
| NGC 6652 | 8.55 | 12.6 | 35.51 | 4.4 | 14 | 14 | 38.21 | 13.56 |
| NGC 6681 | 8.83 | 12.6 | 18.64 | 2.5 | 4 | 4 | 8.61 | 5.82 |
| NGC 6717 | 8.26 | 12.6 | 69.24 | 6.6 | 9 | 10 | 72.48 | 13.19 |
| NGC 6723 | 9.3 | 12.6 | 6.31 | 2.2 | 15 | 14 | 3.94 | 7.86 |
| NGC 6934 | 9.07 | 12.6 | 10.72 | 2.5 | 17 | 18 | 8.61 | 10.12 |
| NGC 6981 | 9.2 | 12.6 | 7.95 | 2.4 | 19 | 18 | 7.06 | 11.18 |
| NGC 7099 | 8.95 | 12.6 | 14.14 | 2.5 | 10 | 10 | 8.61 | 7.68 |
| Median | | | | 2.9 | 14 | 14 | 14.84 | 10.55 |

Table 4c

Transform $f_{2T}= \varpi_2/0.0652- 34.0782$ fit $R^2 = 0.95$

| Cluster name | logtrlx RB | age | $f_{2T}$ | $\varpi_2$ | NBS cal2 | NBS obs TOT | $f_{2T}$ | age2T |
|---|---|---|---|---|---|---|---|---|
| NGC 1904 | 9.1 | 12.6 | 10.01 | 2.7 | 25 | 25 | 7.33 | 9.23 |
| NGC 4372 | 9.59 | 12.6 | 3.24 | 2.3 | 11 | 11 | 1.20 | 4.66 |
| NGC 4590 | 9.29 | 12.6 | 6.46 | 2.6 | 25 | 24 | 5.80 | 11.31 |
| NGC 4833 | 9.34 | 12.6 | 5.76 | 2.4 | 19 | 20 | 2.73 | 5.98 |
| NGC 5927 | 8.98 | 12.6 | 13.19 | 3.1 | 33 | 33 | 13.47 | 12.86 |
| NGC 5986 | 9.23 | 12.6 | 7.42 | 2.5 | 22 | 21 | 4.27 | 7.24 |
| NGC 6093 | 8.86 | 12.6 | 17.39 | 3.3 | 27 | 27 | 16.54 | 11.98 |
| NGC 6218 | 9.02 | 12.6 | 12.03 | 2.8 | 24 | 26 | 8.87 | 9.28 |
| NGC 6266 | 9.19 | 12.6 | 8.14 | 2.6 | 25 | 24 | 5.80 | 8.98 |
| NGC 6356 | 9.26 | 12.6 | 6.92 | 2.5 | 22 | 23 | 4.27 | 7.76 |
| NGC 6362 | 9.31 | 12.6 | 6.17 | 2.5 | 22 | 23 | 4.27 | 8.71 |
| NGC 6638 | 8.51 | 12.6 | 38.94 | 4.8 | 18 | 18 | 39.54 | 12.80 |
| NGC 6637 | 8.79 | 12.6 | 20.43 | 3.4 | 21 | 20 | 18.07 | 11.14 |
| NGC 6642 | 8.49 | 12.6 | 40.77 | 4.8 | 14 | 14 | 39.54 | 12.22 |
| NGC 6838 | 8.41 | 12.6 | 49.02 | 5.5 | 16 | 17 | 50.28 | 12.92 |
| NGC 6864 | 9.08 | 12.6 | 10.48 | 2.8 | 28 | 27 | 8.87 | 10.66 |
| Median | | | | 2.75 | 22 | 23 | 8.10 | 9.97 |

Table 4d

| Cluster name | logtrlx RB | age | $f_{2T}$ | $\varpi_2$ | NBS cal2 | NBS obs TOT | $f_{2T}$ | age2T |
|---|---|---|---|---|---|---|---|---|
| NGC 362 | 8.92 | 12.6 | 15.15 | 3.4 | 41 | 41 | 18.07 | 15.03 |
| NGC 2808 | 9.11 | 12.6 | 9.78 | 3.2 | 46 | 47 | 15.00 | 19.33 |
| NGC 5634 | 9.28 | 12.6 | 6.61 | 2.7 | 27 | 27 | 7.33 | 13.97 |
| NGC 5824 | 9.33 | 12.6 | 5.89 | 3 | 33 | 34 | 11.93 | 25.51 |
| NGC 5904 | 9.53 | 12.6 | 3.72 | 2.5 | 15 | 16 | 4.27 | 14.45 |
| NGC 6229 | 9.19 | 12.6 | 8.14 | 3 | 38 | 38 | 11.93 | 18.48 |
| NGC 6273 | 9.34 | 12.6 | 5.76 | 3 | 33 | 32 | 11.93 | 26.11 |
| Median | | | | 3 | 33 | 34 | 11.93 | 18.48 |

Table 4e

| Cluster name | logtrlx RB | age | $f_{2T}$ | $\varpi_2$ | NBS cal2 | NBS obs TOT | $f_{2T}$ | age2T |
|---|---|---|---|---|---|---|---|---|
| NGC 1261 | 9.2 | 12.6 | 7.95 | 1.8 | 6 | 6 | -2.29 | -3.629 |
| NGC 5946 | 8.95 | 12.6 | 14.14 | 1.8 | 1 | 1 | -2.29 | -2.041 |
| NGC 6284 | 9.16 | 12.6 | 8.72 | 1.8 | 5 | 4 | -2.29 | -3.31 |
| NGC 7078 | 9.35 | 12.6 | 5.63 | 1.9 | 10 | 10 | -0.732 | -1.639 |
| NGC 7089 | 9.32 | 12.6 | 6.03 | 1.6 | 5 | 5 | -5.405 | -11.29 |
| Median | | | | 1.8 | 5 | 5 | -2.29 | -3.31 |